\documentclass[11pt]{article}
\usepackage{amsmath}
\usepackage{amssymb}
\usepackage{amsthm,amsxtra}
\usepackage{epsf}
\usepackage{eepic}
\newlength{\mytopmargin}
\newlength{\myleftmargin}
\newtheorem{cor}{Corollary}

\newtheorem{prop}{Proposition}
\begin{document}
\vspace{4cm}
\noindent
{\bf \Large Matrix averages relating to the Ginibre ensembles} 

\vspace{5mm}
\noindent
Peter J.~Forrester${}^*$
 and Eric M.~Rains${}^\dagger$

\noindent
${}^*$Department of Mathematics and Statistics,
University of Melbourne, \\
Victoria 3010, Australia ; \\
${}^\dagger$
Department of Mathematics, California Institute of Technology, Pasadena, CA 91125, USA

\small
\begin{quote}
The theory of zonal polynomials is used to compute the average of a Schur polynomial of
argument $AX$, where $A$ is a fixed matrix and $X$ is from the real Ginibre ensemble.
This generalizes a recent result of Sommers and Khorozhenko [J. Phys. A {\bf 42} (2009), 222002], 
and furthermore allows
analogous results to be obtained for the complex and real quaternion Ginibre ensembles.
As applications, the positive integer moments of the general variance Ginibre
ensembles are computed in terms of generalized hypergeometric functions, 
these are written in terms of averages over matrices of the same size as the moment
to give duality formulas,
and the averages
of the power sums of the eigenvalues are expressed as finite sums of zonal polynomials.
\end{quote}

\section{Introduction}
The statistical properties of the eigenvalues of random $N \times N$ matrices with
independent, identically distributed (i.i.d.) real entries is a prominent topic in both physics and
mathematics. Physical applications began with the paper of May \cite{Ma72} on the stability of an
ecological network consisting of many components $\{y_i(t)\}_{i=1,\dots,N}$, coupled in some unknown way.
Suppose the evolution of the components is governed by a first order differential system, and
write the linearization of the latter about a fixed point in the form
\begin{equation}\label{1}
\Big [ {d \tilde{y}_i(t) \over d t} \Big ]_{i=1,\dots,N} =
(- \mathbb I_N + B) [ \tilde{y}_i(t)]_{i=1,\dots,N}.
\end{equation}
With the $N \times N$ matrix $B$ having all entries zero this differential equation exhibits exponential
relaxation to the fixed point. For general $B$ the system (\ref{1}) is stable if and only if all the
eigenvalues of $B$ have real part less than or equal to 1.

Since the coupling between components is unknown, it is reasonable to take the components of $B$ to be random.
May argued that for $B$ a dilute matrix (fraction $1-c$ of its elements zero) with non-zero elements i.i.d.~random
variables having mean zero and variance $\sigma^2$, the spectral radius will be less than 1 provided
$\sigma \sqrt{Nc} < 1$.  This is an asympototic result, requiring that $N$ be large. It is consistent with a rigorous
result proved subsequently for the special case $c=1$ \cite{Ge86}.

In the mathematics literature, attention has focussed not only on the spectral radius, but also
the eigenvalue density. With the elements standard Gaussians, by explicit calculation
the eigenvalue density was proved to be asymptotically uniform in the disk of radius $\sqrt{N}$,
centred about the origin in the complex plane \cite{Ed95}. This is the so called circle law \cite{Gi84,Ba97},
which has recently been proved to remain true for general i.i.d.~distributions \cite{TV08}. Another mathematical
property proved true in the Gaussian case by explicit calculation \cite{EKS94} is that for large $N$ the expected number 
of real eigenvalues is asymptotically $\sqrt{2N/\pi}$. Numerical evidence presented in \cite{EKS94}
suggests this result persists for general i.i.d.~distributions of mean zero and unit variance, although a 
proof is yet to be found.

Of interest in both the physics and mathematics literature has been the integrability properties of
eigenvalue distribution in the Gaussian case. Ginibre \cite{Gi65} was the first to seek an analytic
formula for the eigenvalue probability density function (p.d.f.), giving rise to the name real Ginibre
ensemble for real Gaussian matrices. In \cite{Gi65} an analytic expression was found for the
eigenvalue p.d.f.~conditioned so that all eigenvalues are real. For the conditioning specifying a general
number of real eigenvalues, the analytic form of the eigenvalue p.d.f.~was not obtained until the passing
of a further twenty-five years \cite{LS91,Ed95}. And it has not been until the last few years that
analytic computations based on the eigenvalue p.d.f.~have been mastered to the extent that the probability
of a prescribed number of real eigenvalues can be calculated \cite{KA05,AK07,Si07}, and closed form expressions for
the correlations obtained \cite{FN07,Som07,BS07,SW08,FM09}. Furthermore, these analytic studies have been extended
\cite{FN08,APS08} to the case of partially symmetric real Gaussian matrices \cite{LS91}.

A very recent result \cite{SK09} relates to the average of the Schur polynomials $s_\kappa(\lambda_1,
\dots,\lambda_N)$ with respect to the eigenvalues $\{\lambda_j\}$ of matrices from the real Ginibre
ensemble. We recall that the Schur polynomials are the basis for symmetric functions of $\{\lambda_j\}$,
labelled by a partition $\kappa_1 \ge \kappa_2 \ge \cdots \ge \kappa_N \ge 0$ of non-negative
integers, given by the ratio of determinants
\begin{equation}\label{3.0}
s_\kappa(\lambda_1,
\dots,\lambda_N) = {\det [ \lambda_j^{\kappa_k + N - k} ]_{j,k=1,\dots,N} \over
\det [ \lambda_j^{N-k} ]_{j,k=1,\dots,N} }.
\end{equation}
With $\{\lambda_j\}$ being the eigenvalues of the matrix $X$, the Schur polynomials are 
often written $s_\kappa(X)$. Making use of knowledge of the explicit form of the eigenvalue
p.d.f., it is proved in \cite{SK09} that for $X$ a member of the real Ginibre ensemble
\begin{equation}\label{3.1}
\langle s_\kappa(X) \rangle_{X} =
\left \{ \begin{array}{ll} \displaystyle 2^{|\kappa|/2}
\prod_{n=1}^N {\Gamma((N-n+\kappa_n +1)/2) \over \Gamma((N-n+1)/2}, &
{\rm all \:} \kappa_n \: {\rm even} \\
0, & {\rm otherwise} \end{array}
\right.
\end{equation}
where $|\kappa| := \sum_{j=1}^n \kappa_j$.

It is the purpose of this paper to give a different viewpoint on this result. Explicitly, (\ref{3.1}) will be
deduced as a consequence of the theory of zonal polynomials. This viewpoint will allow for
analogues of (\ref{3.1}) to be given for the complex and real quaternion Ginibre ensemble. It further
leads us to the evaluation of the moments of the characteristic polynomial in terms of generalized
hypergeometric functions, in the case that the Gaussian matrices have a (matrix) distribution with
a general variance. This in turn allows us to express the moments as different matrix integrals,
in which the size of the matrix is equal to that of the moments, giving duality formulas. In the
final section, again for the three Ginibre ensembles with a general variance, the averages of
the power sums of the eigenvalues are expressed as finite sums of zonal polynomials.

\section{Zonal polynomials}
\setcounter{equation}{0}
An alternative characterization of the Schur polynomials (\ref{3.1}) is as the eigenfunctions of the
differential operator
\begin{equation}\label{4.0}
\sum_{j=1}^N
\Big (\lambda_j {\partial \over \partial \lambda_j} \Big )^2 +  { N-1 \over \alpha}
\sum_{j=1}^N \lambda_j {\partial \over \partial \lambda_j}
+ {2 \over \alpha}  \sum_{1 \le j < k \le N}{\lambda_j \lambda_k \over \lambda_j -
\lambda_k}
\Big ({\partial \over \partial \lambda_j} -{\partial
\over
\partial \lambda_k}
\Big )
\end{equation}
in the case $\alpha = 1$, with the structure
\begin{equation}\label{4.1}
s_\kappa(\lambda_1,
\dots,\lambda_N) = m_\kappa + \sum_{\mu < \kappa} a_{\kappa \mu} m_\mu.
\end{equation}
In (\ref{4.1}) $m_\kappa$ denotes the monomial symmetric function indexed by the partition $\kappa$
(e.g.~with $N=2$, $m_{1^2} = \lambda_1 \lambda_2 +  \lambda_1 \lambda_3 +  \lambda_2 \lambda_3$),
$\mu < \kappa$ refers to the dominance ordering on partitions specifed by the requirement that
$\sum_{j=1}^l \mu_j \le \sum_{j=1}^l \kappa_j$ $(l=1,\dots,N)$, and the $a_{\kappa \mu}$ are
scalars. For general $\alpha$ the eigenfunctions of (\ref{4.0}) with the structure (\ref{4.1})
are the symmetric Jack polynomials $P_\kappa^{(2/\alpha)}(\lambda_1,\dots,\lambda_N)$.

Associated with a partition $\kappa$ is the generalized Pochhammer symbol 
\begin{equation}\label{5.1}
[u]_\kappa^{(\alpha)} = \prod_{j=1}^N {\Gamma(u - (j-1)/\alpha + \kappa_j) \over
\Gamma(u - (j-1)/\alpha) },
\end{equation}
and in terms of this and the classical Pochhammer symbol
$(u)_n := u(u+1) \cdots (u+n-1)$ one defines
\begin{equation}\label{2.3a}
d_\kappa' = {\alpha^{|\kappa|} [(N-1)/\alpha + 1]_\kappa^{(\alpha)} \over
\bar{f}^{1/\alpha}(\kappa) }, \qquad
\bar{f}^{1/\alpha}(\kappa) := \prod_{1 \le i < j \le N}
{(1 + (j-i-1)/\alpha + \kappa_i - \kappa_j)_{1/\alpha} \over
(1 + (j-i-1)/\alpha)_{1/\alpha}}.
\end{equation}
The quantity $d_\kappa'$ in turn is used to define the renormalized Jack polynomials
\begin{equation}\label{2.4a}
C_\kappa^{(\alpha)}(\lambda_1,\dots,\lambda_N) = {\alpha^{|\kappa|} |\kappa|! \over d_\kappa'}
P_\kappa^{(\alpha)}(\lambda_1,\dots,\lambda_N).
\end{equation}
In the cases $\alpha = 2$, 1 and $1/2$ the renormalized Jack polynomials are the so called zonal polynomials
associated with the symmetric spaces $gl(N,{\mathbb R})/O(N)$,
$gl(N,{\mathbb C})/U(N)$ and $u^*(2N)/Sp(2N)$
\cite{Ma95}.

For present purposes, a key propery of the zonal polynomials is their appearance on the right hand sides of
the matrix integrals
\cite{Ja64, Ma95, Ra95}.
\begin{eqnarray}
\langle s_{\lambda}(A  O) \rangle_{{O} \in O(N)} & = &
\left \{ \begin{array}{ll}
\displaystyle
{C_\kappa^{(2)}(A A^T) \over C_\kappa^{(2)}((1)^N)}, & \lambda = 2 \kappa \\
0, & {\rm otherwise} \end{array} \right.
\label{16.9.1} \\
\langle s_{\lambda}(A U) s_\kappa(U^\dagger A^\dagger)
\rangle_{{U} \in U(N)} & = & \delta_{\lambda, \kappa}
{C_\kappa^{(1)}(A A^\dagger) \over C_\kappa^{(1)}((1)^N)}
\label{16.9.2} \\
\langle s_{\lambda}(A S) \rangle_{{S} \in Sp(2N)} & = &
\left \{ \begin{array}{ll} \displaystyle
{C_\kappa^{(1/2)}(A A^\dagger) \over C_\kappa^{(1/2)}((1)^N)}, &
\lambda = \kappa^2,
\\ 0, & {\rm otherwise} \end{array} \right.
\label{16.9.3}
\end{eqnarray}
where in (\ref{16.9.1}) the partition $2 \kappa$ is the partition obtained
by doubling each part of $\kappa$, while in (\ref{16.9.3}),
$\kappa^2$ is the partition obtained by repeating each part of
$\kappa$ twice. On the left hand sides the averages are over the classical groups
$O(N)$, $U(N)$, $Sp(2N)$ of unitary matrices with real, complex and real quaternion
elements respectively, endowered with the corresponding Haar measure.

We immediately observe a structual similarity between (\ref{3.1}) and (\ref{16.9.1}).
In fact, as will be shown in the next section, (\ref{16.9.1}) implies and furthermore
generalizes (\ref{3.1}).

\section{Averages over the Ginibre ensembles}\label{s3}
\setcounter{equation}{0}
That zonal polynomials are intimately related to averages over matrices with Gaussian
entries is the theme of the monograph by Takemura \cite{Ta84}. This theme has further been developed in
the works \cite{HSS92,Ra95}. More generally, zonal polynomials can be related to any
measure $d\mu(X)$ on the space of random matrices with real, complex or real quaternion
entries possessing the property of being invariant under the mappings $X \mapsto UX$,
$X \mapsto XU$ for $U \in O(N), \, U(N), \, Sp(2N)$ respectively. With $\langle \cdot \rangle_{X}$
denoting an average in such a setting, and $\langle \cdot \rangle_U$ denoting the average over
the corresponding classical group, to apply this theory to the Ginibre ensembles
we first make note of a fundamental factorization property
of the zonal polynomials with respect to the former, namely (see e.g.~\cite{Ma95})
\begin{equation}\label{148}
\langle C_\kappa^{(\alpha)}(A U^\dagger  B  U)
\rangle_U = {C_\kappa^{(\alpha)}(a_1,\dots,a_N) C_\kappa^{(\alpha)}(b_1,\dots,b_N) \over
C_\kappa^{(\alpha)}((1)^N)}.
\end{equation}
Here $C_\kappa^{(\alpha)}((1)^N) := C_\kappa^{(\alpha)}(x_1,\dots,x_N) |_{x_1= \cdots = x_N = 1}$ and
$\{a_i\}$, $\{b_i\}$ are the eigenvalues of $A$, $B$ respectively.

\begin{prop}  \cite{Ta84}
One has
\begin{equation}\label{165}
\langle C_\kappa^{(\alpha)}(A X B X^\dagger) \rangle_X =
{C_\kappa^{(\alpha)}(A) C_\kappa^{(\alpha)}(B) \over (C_\kappa^{(\alpha)}((1)^N))^2 }
\langle C_\kappa^{(\alpha)}(X X^\dagger) \rangle_X.
\end{equation}
\end{prop}

\noindent
{\it Proof.} \quad 
For any $f(X)$ integrable with respect to $d \mu(X)$, the invariance of $d \mu(X)$ under 
$X \mapsto U X$ tells us that
\begin{equation}\label{163}
\langle f(A X B X^\dagger) \rangle_{X}=
\langle \langle f(A U X B  X^\dagger  U^\dagger)
\rangle_{U} \rangle_{X},
\end{equation}
while the invariance under $X \mapsto X U$ gives
\begin{equation}\label{164}
\langle f(A X B X^\dagger) \rangle_{X}=
\langle \langle f(A  X U B U^\dagger X^\dagger)
\rangle_{U} \rangle_{X}.
\end{equation}
Choosing $f = C_\kappa^{(\alpha)}$ in (\ref{163}) and using (\ref{148}) shows
\begin{equation}\label{164c}
\langle C_\kappa^{(\alpha)}(A X B X^\dagger) \rangle_X =
{C_\kappa^{(\alpha)}(A) \over C_\kappa^{(\alpha)}((1)^N)}
\langle  C_\kappa^{(\alpha)}(X B X^\dagger) \rangle_X.
\end{equation}
Choosing  $f = C_\kappa^{(\alpha)}$ in (\ref{164}) and again using (\ref{148}) allows the
right hand side of (\ref{164c}) to be evaluated, and (\ref{165}) results. \hfill $\square$

\medskip
We can use (\ref{165}) combined with (\ref{16.9.1})--(\ref{16.9.3}) to obtain a generalization of
(\ref{3.1}) for each of the real, complex and real Ginibre ensembles. First we consider the case of a
general measure $d \mu(X)$ invariant under multiplication by unitary matrices.

\begin{prop}\label{p1}
Let $d \mu(X)$ be as required for (\ref{165}). For real matrices
\begin{equation}\label{9.1a}
\langle s_\mu(A X) \rangle_{X} =
\left \{ \begin{array}{ll} \displaystyle
{C_\kappa^{(2)}(A  A^T) \over
(C_\kappa^{(2)}((1)^N))^2}\langle C_\kappa^{(1)}(X X^T) \rangle_X, &  \mu = 2 \kappa \\
0, & {\rm otherwise}; \end{array} \right.
\end{equation}
for complex matrices
\begin{equation}\label{9.1b}
\langle s_\mu(A X) s_\kappa(X^\dagger A^\dagger) \rangle_{X}
= \delta_{\lambda, \kappa}
{C_\kappa^{(1)}(A  A^\dagger) \over                    
(C_\kappa^{(1)}((1)^N))^2}
\langle C_\kappa^{(1)}(X X^\dagger) \rangle_X;
\end{equation}
for real quaternion matrices
\begin{equation}\label{9.1c}
\langle s_\mu(A X) \rangle_{X} =
\left \{ \begin{array}{ll} \displaystyle
{C_\kappa^{(1/2)}(A A^T) \over                    
(C_\kappa^{(1/2)}((1)^N))^2} \langle C_\kappa^{(1)}(X X^\dagger) \rangle_X, &  \mu = \kappa^2 \\
0, & {\rm otherwise}. \end{array} \right.
\end{equation}
\end{prop}

\noindent
{\it Proof.} \quad
Consider first (\ref{9.1a}). It follows from (\ref{16.9.1}) that
\begin{align*}
\langle s_\mu(AX) \rangle_X & = \langle \langle s_\mu(A X O) \rangle_O \rangle_X \\
& = \left \{ \begin{array}{ll}\displaystyle {\langle C_\kappa^{(2)}(A^T A X X^T) \rangle_X \over
C_\kappa^{(2)}((1)^N) }, & \mu= \kappa^2 \\
0, & {\rm otherwise}. \end{array} \right.
\end{align*}
Making use of (\ref{165}) in the case $\alpha = 2$ gives (\ref{9.1a}). The derivation of (\ref{9.1b})
and (\ref{9.1c}) begins with (\ref{16.9.2}) and (\ref{16.9.3}), then makes use of (\ref{165}) in an
analogous manner. \hfill $\square$

\medskip
We see from (\ref{9.1a}) and (\ref{5.1}) that (\ref{3.1}) is reclaimed if we can show that for
$d \mu(X) \propto e^{-(1/2) {\rm Tr} \, X X^T} (d X)$,
\begin{equation}\label{11}
{1 \over C^{(2)}_\kappa((1)^N)} \langle C_\kappa^{(2)}(X X^T) \rangle_X = 2^{|\kappa|/2}
[N/2]_\kappa^{(2)}.
\end{equation}
For this we change variables $X X^T = A$, using the result $(d X) \propto (\det A)^{-1/2} (dA)$
(see e.g.~\cite[eq.~(3.30)]{Fo02}),
to obtain
\begin{equation}\label{11a}
\langle C_\kappa^{(2)} (X X^T) \rangle_X = {1 \over C} \int_{A > 0}
e^{-(1/2) {\rm Tr} A} (\det A)^{-1/2} C_\kappa^{(2)}(A) \, (d A)
\end{equation}
where $C$ is such that the RHS equals unity when $\kappa = 0^N$, and $A > 0$ denotes that the
integral is over the space of positive definite matrices. Changing variables now to the eigenvalues and
eigenvectors of $A$ shows that (\ref{11a}) is proportional to
\begin{equation}\label{12}
\int_0^\infty d \lambda_1 \cdots \int_0^\infty d \lambda_N \, \prod_{l=1}^N \lambda_l^{-1/2} e^{- \lambda_l/2}
\, C^{(2)}_\kappa(\lambda_1,\dots,\lambda_N)  \prod_{j<k}^N | \lambda_k - \lambda_j|.
\end{equation}
We recognize the integral in (\ref{12}) as appearing in the integration formula
(see \cite[eq.~(12.152)]{Fo02})
\begin{eqnarray}\label{12a}
&&{1 \over C} \int_0^\infty d \lambda_1 \cdots \int_0^\infty d \lambda_N \, \prod_{l=1}^N \lambda_l^a e^{- \lambda_l}
C_\kappa^{(\alpha)}(\lambda_1,\dots,\lambda_N) \prod_{j<k}^N | \lambda_k - \lambda_j|^{2/\alpha} \nonumber \\
&& \qquad = C_\kappa^{(\alpha)}((1)^N) [a + (N-1)/\alpha + 1]_\kappa^{(\alpha)}
\end{eqnarray}
(take $\alpha = 2$, $a = - 1/2$ and change variables $t_l \mapsto t_l/2$), which itself is a limiting case
of a generalization of the Selberg integral conjectured by Macdonald and proved by
Kadell and Kaneko (see \cite{FW07p} and references therein), and (\ref{11}) follows. 

Proceeding similarly, making use of (\ref{12a}) for $\alpha = 1$, we can show that for
$X$ complex, with $d \mu(X) \propto e^{- {\rm Tr}( X X^\dagger)} (d X)$,
\begin{equation}\label{11d}
{1 \over C_\kappa^{(1)}((1)^N)} \langle C_\kappa^{(1)}(X X^\dagger) \rangle_X =
[N]_\kappa^{(1)}.
\end{equation}
And for $X$ real quaternion with $d \mu(X) \propto e^{- {\rm Tr}( X X^\dagger)} (d X)$ --- the trace now being with
respect to the quaternion structure and so selecting only one of the diagonal elements from each
$2 \times 2$ block --- we have
\begin{equation}\label{11e}
{1 \over C_\kappa^{(1/2)}((1)^N)} \langle C_\kappa^{(1/2)}(X X^\dagger) \rangle_X =
2^{-|\kappa|} [2N]_\kappa^{(2)}.
\end{equation}
Substituting (\ref{11a}), (\ref{11d}) and (\ref{11e}) in Proposition \ref{p1} gives the sought
generalization of (\ref{3.1}) for the Ginibre ensembles.

\begin{cor}\label{c1}
With the distribution of the real Ginibre ensemble proportional to $ e^{-(1/2) {\rm Tr} \, X X^T}$, and
the distribution of the complex and real quaternion ensembles proportional to
$ e^{- {\rm Tr}( X X^\dagger)}$, one has for the real Ginibre ensemble
\begin{equation}\label{13a}
\langle s_\mu(A X) \rangle_{X} =
\left \{ \begin{array}{ll}
\displaystyle   
{2^{|\kappa|} [N/2]_\kappa^{(2)} \over C_\kappa^{(2)}((1)^N)} C_\kappa^{(2)}( A A^T), &  \mu = 2 \kappa \\
0, & {\rm otherwise}; \end{array} \right.
\end{equation}
for the complex Ginibre ensemble
\begin{equation}\label{13b}
\langle s_\mu(AX) s_\kappa(X^\dagger A^\dagger) \rangle_X = \delta_{\mu,\kappa} 
{[N]_\kappa^{(1)} \over C_\kappa^{(1)}((1)^N)} C_\kappa^{(1)}( A  A^\dagger);
\end{equation}
and for the real quaternion Ginibre ensemble
\begin{equation}\label{13c}
\langle s_\mu(A X) \rangle_{X} =
\left \{ \begin{array}{ll}
\displaystyle                                             
{2^{-|\kappa|} [2N]_\kappa^{(1/2)} \over C_\kappa^{(1/2)}((1)^N)} C_\kappa^{(1/2)}( A A^\dagger), &  \mu = \kappa^2 \\
0, & {\rm otherwise}. \end{array} \right.
\end{equation}
\end{cor}

The fact that \cite[Prop.~12.23]{Fo02}
\begin{equation}\label{3.17a}
P_\kappa^{(\alpha)}((1)^N) = {\alpha^{|\kappa|} [N/\alpha]_\kappa^{(\alpha)} \over h_\kappa}
\end{equation}
where \cite[Prop.~12.28]{Fo02} $1/h_\kappa$ is the coefficient of $(x_1 + \cdots + x_N)^{|\kappa|}$ in
$P_\kappa^{(\alpha)}(x)$ allows the results of Corollary \ref{c1} to be written
\begin{eqnarray*}
&&\langle s_\mu(A X) \rangle_{X} =
\left \{ \begin{array}{ll}
h_\kappa P_\kappa^{(2)}( A A^T), &  \mu = 2 \kappa \\
0, & {\rm otherwise}; \end{array} \right. \\
&& \langle s_\mu(AX) s_\kappa(X^\dagger A^\dagger) \rangle_X = \delta_{\mu,\kappa}
h_\kappa P_\kappa^{(1/2)}( A  A^\dagger) \\
&&\langle s_\mu(A X) \rangle_{X} =
\left \{ \begin{array}{ll}
h_\kappa P_\kappa^{(1/2)}( A A^\dagger), &  \mu = \kappa^2 \\
0, & {\rm otherwise}. \end{array} \right.
\end{eqnarray*}
respectively. In this form  Corollary \ref{c1} appears in the unpublished manuscript \cite{Ra95} of one of us (EMR).

\section{Hypergeometric functions}
\setcounter{equation}{0}
We know from workings in \cite{BF03} that
\begin{align}\label{15}
{2^{|\kappa|} [N/2]_\kappa^{(2)} \over C_\kappa^{(2)}((1)^N)} & = {1 \over |\kappa|! 2^{|\kappa|} }
d_{2 \kappa}' |_{\alpha = 1} \nonumber \\
{[N]_\kappa^{(1)} \over C_\kappa^{(1)}((1)^N)} & = {1 \over  |\kappa|! } (d_\kappa'  |_{\alpha = 1})^2 \nonumber \\
{2^{-|\kappa|} [2N]_\kappa^{(1/2)} \over C_\kappa^{(1/2)}((1)^N)} & = {1 \over 2^{|\kappa|} |\kappa|!}
d_{\kappa^2}' |_{\alpha = 1},
\end{align}
which we substitute into the results of Corollary \ref{c1} as appropriate.
To see the consequence of this, we recall the definition of the generalized hypergeometric functions
\begin{equation}\label{15.1}
{}_p^{} F_q^{(\alpha)}(a_1,\dots,a_p, b_1,\dots,b_q;x_1,\dots,x_N) :=
\sum_\kappa {1 \over |\kappa| !}
{[a_1]_\kappa^{(\alpha)} \cdots [a_p]_\kappa^{(\alpha)} \over
[b_1]_\kappa^{(\alpha)} \cdots [b_q]_\kappa^{(\alpha)} }
C_\kappa^{(\alpha)}(x_1,\dots,x_N).
\end{equation}
For $p=0$, $q=1$ this is a generalization of the binomial expansion and we have
\cite[eq.~(13.4)]{Fo02}
\begin{equation}\label{15.2}
{}_1^{} F_0^{(\alpha)}(a;x_1,\dots,x_N) = \prod_{j=1}^N(1 - x_j)^{-a}.
\end{equation}
It follows from this that upon multiplying both sides of (\ref{13a}) and (\ref{13c}) by
$[-r]_\mu^{(1)}/d_\mu'|_{\alpha = 1}$, and both sides of (\ref{13b}) by 
$[-r]_\mu^{(1)}[-r]_\kappa^{(1)}/d_\mu'|_{\alpha = 1} d_\kappa'|_{\alpha = 1} $,
and summing over $\mu$ (or $\mu$ and $\kappa$ in the case of (\ref{13b})), the left hand sides of
each of the identities can be summed according to (\ref{15.2}). For the resulting matrix averages to be
well defined, we require $r \in \mathbb Z_{\ge 0}$. To simplify the right hand sides, we note from (\ref{5.1})  that 
$$
[u]_{2 \kappa}^{(1)} = 2^{2 |\kappa|} [u/2]_\kappa^{(2)} [(u+1)/2]_\kappa^{(2)}, \qquad 
[u]_{\kappa^2}^{(1)}= [u]_\kappa^{(1/2)} [u-1]_\kappa^{(1/2)},
$$
then make use of (\ref{15.1}). Consequently
we obtain the following set of 
matrix integral evaluations.

\begin{cor}
Let $r \in \mathbb Z_{\ge 0}$.
One has, for the real, complex and real quaternion  Ginibre ensembles respectively,
\begin{align}\label{16}
& \langle \det (\mathbb I_N - A X)^{r} \rangle_X =
{}_2^{} F_0^{(2)}(-r/2,(-r+1)/2;2 A A^T) \nonumber \\
& \langle \det (\mathbb I_N - A X)^{r} \det(\mathbb I_N - X^\dagger A^\dagger)^{r} \rangle_X =
{}_2^{} F_0^{(1)}(-r,-r; A A^\dagger)  \nonumber \\
&  \langle \det (\mathbb I_{2N} - A X)^{r} \rangle_X = {}_2^{} F_0^{(1/2)}(-r,-r-1; A A^\dagger/2).
\end{align}
\end{cor}

Suppose $A$ in (\ref{16}) is invertible, and write $\Sigma = (A^\dagger A)^{-1}$. We can then rewrite (\ref{16})
to read
\begin{align}\label{16a}
& \langle \det (\mathbb I_N - x X)^{r} \rangle_X =
{}_2^{} F_0^{(2)}(-r/2,(-r+1)/2;2 x^2 \Sigma) \nonumber \\
& \langle \det (\mathbb I_N - x X)^{r} \det(\mathbb I_N - \bar{x} X^\dagger )^{r} \rangle_X =
{}_2^{} F_0^{(1)}(-r,-r; |x|^2 \Sigma )  \nonumber \\
&  \langle \det (\mathbb I_{2N} - x X)^{r} \rangle_X = {}_2^{} F_0^{(1/2)}(-r,-r+1; |x|^2 \Sigma /2).
\end{align}
Here the averages are over $N \times N$ matrices $X$ with real, complex and real quaternion elements
respectively having a distribution proportional to $e^{-{\rm Tr} (X X^T \Sigma^{-1})/2}$ in the real case
and $e^{-{\rm Tr} (X X^\dagger \Sigma^{-1})}$ in the complex and quaternion real cases.
Introducing $Y = X - (1/x) \mathbb I_N$ and setting $W = Y Y^\dagger$ we then have that $W$ is distributed
as a non-central Wishart distribution (see e.g.~\cite{Mu82}) in the first case, and its complex and real
quaternion generalization in the other two cases. Furthermore, for $r=2s$ even in the real 
case, and for general non-negative integer $r$ in the complex and real quaternion
cases, the determinants in (\ref{16a}) can
be written entirely in terms of $W$, and we obtain 
\begin{align}\label{16b}
& |x|^{2Ns} \langle (\det W)^{s} \rangle_W =
{}_2^{} F_0^{(2)}(-s,-s+1/2;2 x^2 \Sigma) \nonumber \\
& |x|^{2Nr} \langle (\det W)^{r} \rangle_W =
{}_2^{} F_0^{(1)}(-r,-r; |x|^2 \Sigma )  \nonumber \\
&  |x|^{Nr} \langle (\det W)^{r/2} \rangle_W = {}_2^{} F_0^{(1/2)}(-r,-r-1; |x|^2 \Sigma /2).
\end{align}

The identities (\ref{16b}) are noteworthy for the fact that in the real case a
different generalized hypergeometric function evaluation is known
\cite[Th.~10.3.7]{Mu82},
\begin{equation}
 \langle (\det W)^{s} \rangle_W = (\det \Sigma)^{s} 2^{Ns} {\Gamma_m((N/2+s)) \over \Gamma_m(N/2)}
\, {}_1^{} F_1^{(2)}(-s;N/2;-\Sigma^{-1}/2x^2)
\end{equation}
where
$$
\Gamma_m(u) := \pi^{m(m-1)/2} \prod_{j=1}^m \Gamma(u - (i-1)/2).
$$
This implies (after some minor simplification) the identity between generalized hypergeometric functions
\begin{equation}\label{id}
{}_2^{} F_0^{(2)}(-s,-s+1/2; Y) = [N/2]_{s^N}^{(2)} (\det Y)^s \, {}_1^{} F_1^{(2)}(-s;N/2;- Y^{-1}).
\end{equation}
Using the property of Jack polynomials \cite[Exercises 12.1 q.1 and 2]{Fo02}
$$
(\det Y)^s P_\kappa^{(\alpha)}(Y^{-1}) = P_{\kappa^s}^{(\alpha)}(Y)
$$
where $\kappa^s := (s - \kappa_N, s- \kappa_{N-1},\dots,s - \kappa_1)$, noting from (\ref{2.3a}) that
$$
{2^{|\kappa|} \over 2^{|\kappa^s|} }
{d_{\kappa^s}' \over d_\kappa'} =
{[(N+1)/2]_{\kappa^s}^{(2)} \over [(N+1)/2]_{\kappa}^{(2)}},
$$
and using the property of the generalized Pochhammer symbol (\ref{5.1})
$$
[u]_{\kappa^s}^{(\alpha)} = (-1)^{|\kappa^s|}{ [(N-1)/\alpha - u + 1 - s]_{s^N}^{(\alpha)} \over
 [(N-1)/\alpha - u + 1 - s]_{\kappa}^{(\alpha)}}
$$
(a consequence of the functional equation for the gamma function), (\ref{id}) can be verified directly
by comparing coefficients of $P_\kappa^{(2)}(Y)$ on both sides. 

\section{Duality identities}
\setcounter{equation}{0}
There are many matrix ensembles of $N \times N$ matrices $\{X\}$ for which 
$\langle \det (\mathbb I_N - x X)^r \rangle_X$ can be expressed in terms of an average over
dual matrix ensembles where the size of the matrices is $r \times r$ (see e.g.~\cite{Fo02}).
An example of relevance to the present study is an identity of Fyodorov and
Khorozhenko \cite{FK06} (see also \cite{FS09}), which reads
\begin{eqnarray}\label{r.1}
&& 
\langle | \det (z \mathbb I_N - A U) |^{2p} \rangle_{U \in U(N)} \nonumber \\
&& \qquad \propto
\int_0^\infty dt_1 \cdots \int_0^\infty dt_p \,
\prod_{l=1}^p {\det (|z|^2 \mathbb I_p + t_l A A^\dagger) \over (1 + t_l)^{N+2p} }
\prod_{1 \le j < k \le p} |t_k - t_j|^2.
\end{eqnarray}
To write the right hand side as a matrix average, we require a result from random matrix
theory \cite[Exercises 3.6 q.3]{Fo02} giving that the matrix $Y = X^\dagger (B^{-1/2})^\dagger
B^{-1/2} X$, where $X$ is an $(N+p) \times p$ standard complex Gaussian matrix, and $A$ is
a complex Wishart matrix $B = b^\dagger b$ with $b$ and $(N+p) \times (N+p)$ standard complex Gaussian 
matrix, has eigenvalue p.d.f.~proportional to
$$
\prod_{l=1}^p {1 \over (1 + t_l)^{N+2p} }
\prod_{1 \le j < k \le p} |t_k - t_j|^2.
$$
Noting too that
$$
\prod_{l=1}^p \det (|z|^2 \mathbb I_p + t_l A A^\dagger) = \det(|z|^2  \mathbb I_{Np} + Y \otimes A A^\dagger)
$$
we see (\ref{r.1}) can be rewritten as the matrix average duality
\begin{equation}\label{r.2}
\langle |\det (z \mathbb I_N - A U) |^{2p} \rangle_{U \in U(N)} =
\langle \det (|z|^2 \mathbb I_{Np} + Y  \otimes A A^\dagger) \rangle_Y.
\end{equation}

The close relationship seen in \S \ref{s3} between averages over the unitary group which contain an
arbitrary matrix, and averages over the complex Ginibre ensemble, suggests analogous formulas hold for the
averages in (\ref{16a}). This is indeed the case. The sought identities follow from a 
representation for ${}_2 F^{(\alpha)}_0(-r,-a/\alpha-(r-1);Y)$, $r \in \mathbb Z^+$ as an
$r$-dimensional integral, valid for all $a > 0$.

\begin{prop}
Let $r \in \mathbb Z^+$, $a>0$ and $Y$ be an $N \times N$ matrix. We have
\begin{eqnarray}\label{r.3}
&& {}_2 F^{(\alpha)}_0(-r,-a/\alpha-(r-1);Y) 
\nonumber \\
&& \qquad
= {1 \over \widetilde{W}_{a-1,2\alpha,r}}
\int_0^\infty dt_1 \cdots \int_0^\infty dt_r \, \prod_{l=1}^r e^{- t_l} t_l^{a-1}
\det (\mathbb I_N + (t_l/\alpha) Y) \prod_{1 \le j < k \le r} |t_k - t_j|^{2\alpha},
\end{eqnarray}
where
$$
\widetilde{W}_{\lambda_1,\beta,n} :=
\int_0^\infty dt_1 \cdots \int_0^\infty dt_n \, \prod_{l=1}^n  e^{- t_l} t_l^{\lambda_1}
 \prod_{1 \le j < k \le n} |t_k - t_j|^\beta
$$
(this normalization is a well known limiting case of the Selberg integral, and as such
can be evaluated as a product of gamma functions \cite[Prop.~4.7.3]{Fo02}, although we don't
need this fact). 
\end{prop} 

\noindent
{\it Proof.} \quad With $\kappa'$ denoting the conjugate partition, obtained by interchanging
the rows and columns of the diagram of $\kappa$, we have
\cite[Exercises 12.4 q.2]{Fo02}
$$
[u]_{\kappa'}^{(\alpha)} = (-\alpha)^{-|\kappa|} [-\alpha u]_\kappa^{(1/\alpha)}.
$$
Also, from the definition of $d_\kappa'$ in terms of arm and leg lengths \cite[Eq.~(12.37)]{Fo02}, and
the corresponding definition of $h_\kappa$  \cite[Eq.~(12.58)]{Fo02}, one sees
$$
d_{\kappa'}' = \alpha^{|\kappa|} h_\kappa |_{\alpha \mapsto 1/\alpha}.
$$
Recalling (\ref{2.4a}) and (\ref{15.1}) it follows
\begin{equation}\label{F8}
{}_2 F_0^{(\alpha)}(-r,-a/\alpha - (r-1); x_1, \dots, x_N) =
\sum_\kappa { \alpha^{-2|\kappa|} [r \alpha]_\kappa^{(1/\alpha)}
[a + (r-1)\alpha]_\kappa^{(1/\alpha)} \over
 h_\kappa |_{\alpha \mapsto 1/\alpha} }
P_{\kappa'}^{(\alpha)}(X).
\end{equation}
The significance of this series form, as distinct from the form implied by (\ref{15.1}), is
that the coefficient of $P_{\kappa'}^{(\alpha)}(X)$ permits the integral representation
\begin{eqnarray*}
 && { \alpha^{-|\kappa|} [r \alpha]_\kappa^{(1/\alpha)}
[a + (r-1)\alpha]_\kappa^{(1/\alpha)} \over
 h_\kappa |_{\alpha \mapsto 1/\alpha} } \nonumber
\\ && \qquad =
{1 \over \widetilde{W}_{a - 1,2\alpha,r} }
\int_0^\infty dt_1 \cdots \int_0^\infty dt_r \, \prod_{l=1}^r t_l^{a-1} e^{-t_l} 
P_\kappa^{(1/\alpha)}(T) \prod_{1 \le j < k \le r} 
|t_k - t_j|^{2 \alpha},
\end{eqnarray*}
which is just a rewrite of (\ref{12a}), after making use of (\ref{3.17a}).
Substituting in (\ref{F8}), and recalling the dual Cauchy identity for Jack polynomials
\cite[Eq.~(12.186)]{Fo02}
$$
\prod_{k,l=1}^N(1 + x_k y_l) = \sum_\kappa P_\kappa^{(\alpha)}(X) P_{\kappa'}^{(1/\alpha)}(Y),
$$
(\ref{r.3}) follows. \hfill $\square$

\begin{cor}
Let the averages over $X$ be as in (\ref{16a}). We have
\begin{eqnarray}
&& \langle \det (x \mathbb I_N - X)^{2s} \rangle_X \nonumber \\
&& \qquad = {1 \over \widetilde{W}_{0,4,r} }
\int_0^\infty dt_1 \cdots \int_0^\infty dt_s \, \prod_{l=1}^s e^{- t_l}
\det (x^2 \mathbb I_N + t_l \Sigma) \prod_{1 \le j < k \le s} (t_k - t_j)^4 \label{N1} \\
&& \langle |\det (x \mathbb I_N - X) |^{2r} \rangle_X \nonumber \\
&& \qquad = {1 \over \widetilde{W}_{0,2,r} }
\int_0^\infty dt_1 \cdots \int_0^\infty dt_r \, \prod_{l=1}^r e^{- t_l}
\det (|x|^2 \mathbb I_N + t_l \Sigma) \prod_{1 \le j < k \le r} (t_k - t_j)^2 \label{N2} \\
&& \langle \det (x \mathbb I_{2N} - X)^{r} \rangle_X \nonumber \\
&& \qquad = {1 \over \widetilde{W}_{0,1,r} }
\int_0^\infty dt_1 \cdots \int_0^\infty dt_r \, \prod_{l=1}^r e^{- t_l}
\det (|x|^2 \mathbb I_{2N} + t_l \Sigma) \prod_{1 \le j < k \le r} |t_k - t_j|. \label{N3}
\end{eqnarray}
\end{cor}

Each of the right hand sides in the above identities can be written as matrix
averages involving Wishart matrices. For example, with $Y = a^\dagger a$, where $a$
is an $r \times r$ matrix of standard complex Gaussian entries, (\ref{N2}) can be written
\begin{equation}
\langle | \det (x \mathbb I_N - X) |^{2r} \rangle_X =
\langle  \det (|x|^2 \mathbb I_{2Nr} + Y \otimes \Sigma)  \rangle_Y.
\end{equation}

In the theory of the complex Ginibre ensemble with $\Sigma = \mathbb I_N$ it is well known that the
eigenvalue density $\rho_{(1)}((x,y))$ (with $N \mapsto N + 1$ for convenience) satisfies
\begin{equation}\label{X1}
\rho_{(1)}((x,y)) = {e^{-|z|^2} \over \pi N!} \langle  | \det (z \mathbb I_N - X) |^{2} \rangle_X
\end{equation}
(see e.g.~\cite[eq.~(1.3)]{FK06}). Here $z = x+iy$ and $\{X\}$ is the complex Ginibre ensemble with
$\sigma = \mathbb I_N$. Even though the equality is restricted to $\Sigma = \mathbb I_N$, this
motivates us to use the duality identity (\ref{N2}) to
investige the right hand side of (\ref{X1}) when $\{X\}$ is the complex Ginibre
ensemble for a more general variance matrix. In particular, suppose
$\Sigma = {\rm diag} \,(\sigma, (1)^{N-1})$. Then (\ref{N2}) gives
that the right hand side of (\ref{X1}) equals
\begin{equation}\label{RHS}
{\sigma \over \pi N!} \int_{|z|^2}^\infty e^{-t} t^N \, dt +
{(1-\sigma) |z|^2 \over \pi N!}  \int_{|z|^2}^\infty e^{-t} t^{N-1} \, dt.
\end{equation}
For $0 < |z|^2 < N$ and $N$ large the leading behaviour is
$$
{\sigma \over \pi} + {(1 - \sigma) |z|^2 \over \pi N}
$$
while for $|z| = \sqrt{N} - r$ and $\sigma$ fixed, for large $N$ we obtain
$$
{1 \over 2 \pi} (1 + {\rm erf}(\sqrt{2} r) )
$$
independent of $\sigma$. Since the boundary of the eigenvalue support is $|z|= \sqrt{N}$,
this suggests our choice of $\Sigma$ did not effect the largest eigenvalues.
On the other hand, if we were to set $\sigma \mapsto \sigma \sqrt{N}$, as well
as $|z| = \sqrt{N} - r$, the asymptotic form of (\ref{RHS}) is dependent on
$\sigma$, indicating that then the
largest eigenvalues
have been altered.

\section{Power sum averages}
\setcounter{equation}{0}
In the previous section we computed, in (\ref{16a}), the integer moments of the characteristic
polynomials for real, complex and real quaternion Gaussian matrices with a general variance matrix
$\Sigma$. Here we average the power sums $p_k(X) := {\rm Tr} \,(X^k)$ over the same class of
matrices. For real Gaussian matrices
with $\Sigma = \mathbb I_N$, this average was computed using (\ref{13a}) in the case $A = \mathbb I_N$
\cite{SK09}.
If we make use of each of the identities of Corollary \ref{c1} for general $A$, together with the
expansion \cite[Ex.~1.4.10]{Ma95}
$$
p_k(X) = \sum_{l=0}^{k-1} (-1)^l s_{(k-l,1^l)}(X)
$$
we can generalize the result of \cite{SK09}.

\begin{cor}\label{c4}
Let the averages over $X$ be as in (\ref{16a}).
We have
\begin{eqnarray*}
&& \langle p_k(X) \rangle_X = \left \{ \begin{array}{ll}0, & k \: \: {\rm odd} \\
\displaystyle 2^{k/2} \sum_{l=0}^{k/2 - 1} {[N/2]^{(2)}_{(k/2-l,1^l)} \over C_{(k/2-l,1^l)}^{(2)}((1)^N)}
 C_{(k/2-l,1^l)}^{(2)}(\Sigma), &  k \: \: {\rm even} \end{array} \right. \\
  &&\langle p_k(X) p_k(X^\dagger) \rangle_X = \sum_{l=0}^{k-1}
{[N]_{(k-l,1^l)}^{(1)} \over  C_{(k-l,1^l)}^{(1)}((1)^N)}  C_{(k-l,1^l)}^{(1)}(\Sigma) \\ 
&&\langle p_k(X) \rangle_X = \left \{ \begin{array}{ll}0, & k \: \: {\rm odd} \: \: {\rm or} \: \: k/2 > N\\
\displaystyle 2^{-k/2} {[2N]^{(1/2)}_{1^{k/2}} \over C_{1^{k/2}}^{(1/2)}((1)^N)}
  C_{1^{k/2}}^{(1/2)}(\Sigma), &  k \le 2N \: \: {\rm even}. \end{array} \right.
\end{eqnarray*}
\end{cor}

Let $\rho_{(1)}^{\rm r}$ and $\rho_{(1)}^{\rm c}((x,y))$ denote the density of the real and (upper half
plane) complex eigenvalues for real Gaussian matrices with variance matrix $\Sigma$.
Then we have
$$
\langle p_k(X) \rangle_X = \int_{-\infty}^\infty x^k \rho_{(1)}^{\rm r}(x) \, dx +
\int_{\mathbb R_+^2}((x+iy)^k + (x-iy)^k) \rho_{(1)}^{\rm c}((x,y)) \, dx dy
$$
where $\mathbb R_+^2 := \{ (x,y): \, x \in \mathbb R, \, y \in \mathbb R^+ \}$. The average in
Corollary \ref{c4} in the complex case can be rewritten as an average involving both the 
one and two point correlations, which we refrain from writing down. In the real quaternion case, the average
again can be written in terms of just the one point density. Here there are no real eigenvalues, and
we have
$$
\langle p_k(X) \rangle_X = \int_{\mathbb R_+^2}((x+iy)^k + (x-iy)^k) \rho_{(1)}^{\rm c}((x,y)) \, dx dy.
$$

\section*{Acknowledgements}
The work of PJF was supported by the Australian Research Council.

\providecommand{\bysame}{\leavevmode\hbox to3em{\hrulefill}\thinspace}
\providecommand{\MR}{\relax\ifhmode\unskip\space\fi MR }
\providecommand{\MRhref}[2]{%
  \href{http://www.ams.org/mathscinet-getitem?mr=#1}{#2}
}
\providecommand{\href}[2]{#2}

%\bibliographystyle{amsplain}
%\bibliography{book1}

\end{document}